\documentclass[10pt,conference]{IEEEtran}
\IEEEoverridecommandlockouts
\usepackage{cite}
\usepackage{amsmath,amssymb,amsfonts}
\usepackage{algorithmic}
\usepackage{booktabs}
\usepackage{graphicx}
\usepackage{multirow}
\usepackage{textcomp}
\usepackage{xcolor}
\usepackage{optidef}
\usepackage{subcaption}
\usepackage{xspace}
\usepackage[hyphens,spaces,obeyspaces]{url}
\def\BibTeX{{\rm B\kern-.05em{\sc i\kern-.025em b}\kern-.08em
    T\kern-.1667em\lower.7ex\hbox{E}\kern-.125emX}}
    
\begin{document}

\title{An Energy-Aware Approach to Design Self-Adaptive AI-based Applications on the Edge}

\makeatletter
\newcommand{\linebreakand}{%
  \end{@IEEEauthorhalign}
  \hfill\mbox{}\par
  \mbox{}\hfill\begin{@IEEEauthorhalign}
}
\makeatother



\author{
\IEEEauthorblockN{
Alessandro Tundo\IEEEauthorrefmark{1}, 
Marco Mobilio\IEEEauthorrefmark{1}, 
Shashikant Ilager\IEEEauthorrefmark{2} 
\linebreakand
Ivona Brandić\IEEEauthorrefmark{2}, 
Ezio Bartocci\IEEEauthorrefmark{2}, 
Leonardo Mariani\IEEEauthorrefmark{1}}
\linebreakand\\
\IEEEauthorblockA{\IEEEauthorrefmark{1}\textit{University of Milano-Bicocca}\\
Milan, Italy \\
\{name.surname\}@unimib.it}
\and\\
\IEEEauthorblockA{\IEEEauthorrefmark{2}\textit{Vienna University of Technology}\\
Vienna, Austria \\
\{name.surname\}@tuwien.ac.at}
}

\maketitle

\newcommand{\nopedestriansmode}{power-saving\xspace}
\newcommand{\fewpedestriansmode}{low-energy\xspace}
\newcommand{\smallgroupmode}{high-accuracy\xspace}
\newcommand{\crowdmode}{high-rate\xspace}

\newcommand{\setblack}{\color{black}}
\newcommand{\revised}[2]{#2\xspace}
\newcommand{\setred}{}

\newcommand{\repo}{\url{https://gitlab.com/sustainable-continuum-monitoring/self-adaptive-moop/-/tree/ASE_2023?ref_type=tags}\xspace}

\begin{abstract}
The advent of edge devices dedicated to machine learning tasks enabled the execution of AI-based applications that efficiently process and classify the data acquired by the resource-constrained devices populating the Internet of Things. The proliferation of such applications (e.g., critical monitoring in smart cities) demands new strategies to make these systems also sustainable from an energetic point of view.  

In this paper, we present an energy-aware approach for the design and deployment of self-adaptive AI-based applications that can balance application objectives (e.g., accuracy in object detection and frames processing rate) with energy consumption. We address the problem of determining  the set of configurations that can be used to self-adapt the system with a meta-heuristic search procedure that only needs a small number of empirical samples. The final set of configurations are selected using weighted gray relational analysis, and mapped to the operation modes of the self-adaptive application.
 
We validate our approach on an AI-based application for pedestrian detection.
Results show that our self-adaptive application can outperform non-adaptive baseline configurations by saving up to 81\% of energy while loosing only between 2\% and 6\% in accuracy.

\end{abstract}

\begin{IEEEkeywords}
self-adaptive, energy-aware, AI-based, multi-objective, edge computing, internet-of-things
\end{IEEEkeywords}

 
\section{Introduction}
\label{sec:introduction}

Both academia and industry raised the issue of the massive amount of energy consumed by ICT services and the rising energy costs
~\cite{iea2022data,uk2022energy,lange2020digitalization,fonseca2019manifesto}.
Reducing energy consumption is a high priority objective, to wisely use the available resources. Indeed, building \emph{sustainable AI-based applications} is a key technical challenge that engineers are facing nowadays~\cite{jiang2020energy}. 

AI-based applications are increasingly deployed on the \emph{edge}, within resource-constrained environments that cannot indefinitely supply a constant amount of power, such as, battery-powered devices and computing nodes powered by renewable energy sources (e.g., photovoltaic panels or wind turbines)~\cite{callebaut2021art,elahi2020energy,pecunia2021emerging}. \revised{MR.C3, R1.C2}{Such applications are particularly resource-intensive, thus, carefully} using energy is a key requirement to feasibly run AI services within these environments. For example, critical monitoring services for smart cities (e.g., pedestrian detection and traffic analysis~\cite{lujic2021increasing,combs2019automated,nagy2018survey}), environmental monitoring applications (e.g., wildfire detection~\cite{lin2018kalman,allison2016airborne}, and wildlife monitoring~\cite{dominguez2021wildlife,schwartz2021deployment}), all require fast data processing and high accuracy, with cost-effective energy consumption.

These scenarios require consuming a large volume of data generated from Internet-of-Things (IoT) sensors in various forms (e.g., time series values, video streams, images) with resource-greedy machine learning models (e.g., exploiting TPUs or GPUs)~\cite{garcia2019estimation,reuther2019survey,jiang2020energy}. In contrast, the feasibility of scenarios that involve battery-powered devices~\cite{abrar2021energy,arouj2022towards} depends on the capability of reducing energy consumption to extend the battery life. 

In response to this urge, researchers have investigated several approaches to design systems with \emph{a controllable and programmable trade-off among quality, efficiency, and energy consumption}.

Energy-awareness and efficiency research mainly targets low-level tasks such as scheduling and provisioning~\cite{aral2021ares,aljulayfi2019novel,ghosh2018adaptive,nan2017adaptive,silva2021energy}, routing~\cite{khan2021monitoring}, data storage and processing~\cite{vales2019energy}, and machine learning models optimization~\cite{brandic2021sustainable}. Although valuable, only optimizing the low-level tasks may result in hardly-predictable performance of the applications. Thus, it becomes challenging or even impossible to balance competing application-level objectives (e.g., accuracy, energy consumption, and efficiency) working only on low-level features.

Other approaches targeted code optimizations~\cite{sahar2019towards}, analysis of software energy consumption~\cite{trefethen2013energy,eder2017energy}, and architectural tactics to contain energy utilization~\cite{chinnappan2021architectural} and costs~\cite{vos2022architectural}.
Analyzing energy consumption retrospectively to take corrective actions (e.g., code or architectural refactoring) can be expensive and difficult to control in the long term.

In contrast with previous work, in this paper we investigate the challenge of configuring (e.g., determining the frame rate, the image resolution, and the kind of machine-learning model that an object detection system must use) AI-based edge applications, to \emph{balance} energy consumption and application objectives. \revised{MR.C3, R1.C2}{Despite this being a common challenge to any edge application, we target AI-based applications since they are frequently used in the edge, despite being resource-demanding.}

Naively, we could hypothesize to simply systematically and exhaustively explore the configuration space of an application, and then determine the best configuration to use. In practice, there are two main obstacles: the \emph{huge cost of the exploration of the configuration space} and the \emph{lack of a configuration that globally optimizes every objective}.

Exploring the configuration space of AI-based edge applications is extremely expensive due to the \emph{size} of the space, determined by the high number of configuration parameters and parameter values, and the \emph{cost of sampling}, which requires running multiple experiments, to determine how much a configuration fulfills the energy and application objectives~\cite{panerati2017optimization}. This cost is even higher in large distributed and heterogeneous environments, where different nodes or groups of nodes may require to be optimized individually.

Furthermore, \emph{different run-time scenarios usually require different configurations} to be addressed properly. For instance, detecting objects in situations where the objects to be detected occur rarely (e.g., detecting pedestrians at night in a peripheral city area) is completely different from detecting the same objects in situations where the objects occur densely and repeatedly (e.g., detecting pedestrians in an area near a stadium after a concert). 
Thus, no single configuration can optimize both accuracy and energy consumption in all circumstances, but applications need to adapt to changing conditions to behave optimally. 

We address these challenges by proposing an \emph{energy-aware} approach that can guide developers to implement an \emph{AI-based self-adaptive application} able of switching its operation modes in response to changes in the environment, finally balancing energy consumption with the application-level objectives.

In a nutshell, this work provides the following contributions.
 
\textbf{An energy-aware approach for the design of AI-based self-adaptive applications.} We present an approach to design and implement an AI-based self-adaptive application that can dynamically balance application requirements and energy consumption, according to a behavioral model derived empirically.

\textbf{A meta-heuristic search procedure combined with a weighted configuration extraction process.} We define a meta-heuristic search procedure that allows to empirically sample a tiny portion of the configuration search space, to finally extract, using \emph{weighted gray relational analysis}, a set of configurations that correspond to the operation modes employed by the self-adaptive system.

\textbf{A smart city scenario prototype implementation.} We showcase the applicability of the proposed approach by implementing the prototype of an AI-based self-adaptive application for a pedestrian detection scenario involving a single-board computer equipped with a camera and a hardware accelerator (i.e., an Edge TPU).

\textbf{Empirical evidence of the effectiveness of the approach.} We answer two research questions by performing in-lab experiments and evaluating pedestrian detection scenarios following real-word pedestrian traffic shapes. Results show that configurations obtained through the meta-heuristic search procedure perform comparably well with respect to the ones obtained by a near-exhaustive search of the space.
The comparison to four non-adaptive baseline applications shows that the self-adaptive system is able to self-adapt its operation mode to the pedestrian traffic shapes saving up to 81\% of energy consumption. At the same time, it guarantees a similar accuracy when compared to the most accurate configurations, losing between 2\% and 6\% only, but outperforming 3 out of 4 non-adaptive applications on the processing speed gaining between 77\% and 233\%.

The paper is organized as follows. Section~\ref{sec:montivational-scenario} presents a Smart Traffic Monitoring (STM) motivational scenario. Section~\ref{sec:approach} describes our approach, with specific reference to the motivational scenario. Section~\ref{sec:evaluation} presents the empirical results. Section~\ref{sec:related-work} discusses related work. Finally, Section~\ref{sec:conclusions} presents concluding remarks and future work.

 
\section{Motivational Scenario}
\label{sec:montivational-scenario}
According to the latest report released by Governors Highway Safety Association (GHSA), nearly 3.500 pedestrians died in the United States in the first six months of 2022 (+5\% from the same period in 2021)~\cite{ghsa2023pedestrians}. In three years, pedestrian deaths raised about 18\%, that is, nine times faster than U.S. population growth~\cite{forbes2023pedestrians}.
Similarly, the European Transport Safety Council (ETSC) reported 20.600 road deaths in the EU last year, with vulnerable road users (pedestrians, cyclists, and users of powered two-wheelers) representing just under 70\% of total fatalities within urban areas~\cite{european2023five,european2023road}. Addressing this critical issue of preventing accidents not only depends on social education~\cite{cdc2022pedestrians} but also requires developing Smart Traffic Monitoring (STM) systems that enable digital monitoring of urban traffic ~\cite{shreyas2017dynamic,alagarsamy2020designing,bolsunovskaya2021development}, real-time analytics ~\cite{amini2017big, barthelemy2019edge}, and intelligent driver assistants~\cite{voinea2020driving,lashkov2020smartphone,lujic2021increasing}.

\begin{figure}[!ht]
    \centering
    \includegraphics[width=\columnwidth]{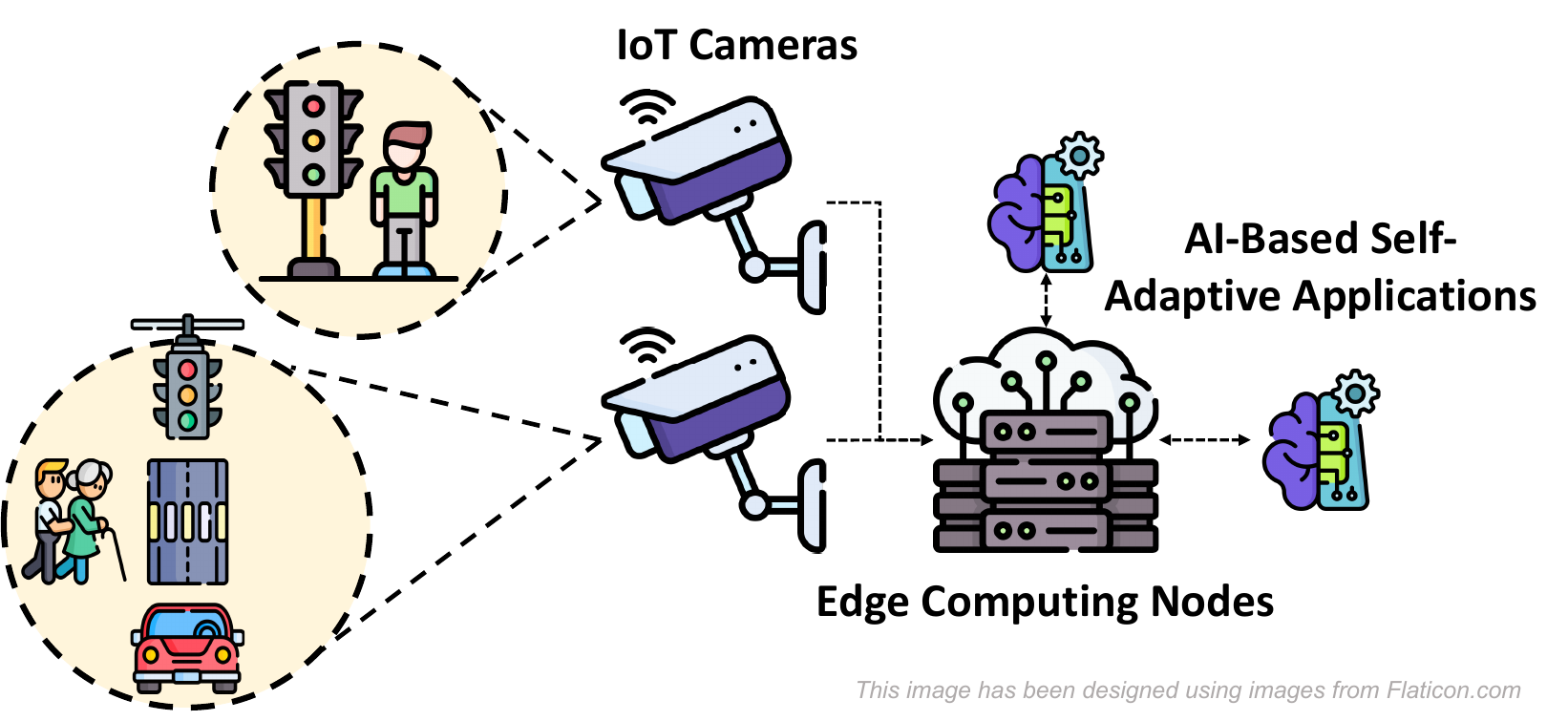}
    \caption{A pedestrian detection scenario.}
    \label{fig:scenario}
\end{figure}

An STM system requires continuous monitoring of the traffic scenarios to identify potential incidents (e.g.,  the presence of pedestrians in blind spots) through video streams and processing frames, and alerting the nearby vehicles through the use of 5G-enabled edge nodes~\cite{lujic2021increasing}. Such an STM system can host hundreds of cameras and sensors deployed to roads in cities and countryside areas ~\cite{dobler2021patterns}.

The edge devices processing video streams are in always-on mode and potentially powered by batteries or renewable energy sources at the edge, which is the basis for limited and unreliable power supply. Hence, reducing energy consumption and executing critical emergency applications become extremely important. On the other hand, such critical applications expect a minimum QoS for safety and reliability (e.g., inference time and ML model accuracy). Therefore, they require continuous monitoring of resources (e.g., energy budget) and workload (e.g., number of detected pedestrians in time intervals), and when needed, employing self-adaptive applications and adapting hardware and software configurations (e.g., camera resolution, ML model, and hardware acceleration).

Figure~\ref{fig:scenario} depicts a pedestrian detection scenario where an application can employ different operation modes according to pedestrian traffic volumes. For instance, this scenario could be addressed with four operation modes as defined in Table~\ref{tab:operation-modes}. 
A self-adaptive application for this scenario can autonomously balance resource (e.g., energy consumption) and application requirements (e.g., frame processing speed and accuracy) by switching among the different operation modes.

On the contrary, using a single operation mode for a whole day cannot adapt to a changing environment. Considering a smart-city scenario with hundreds of IoT cameras and dozens of application instances deployed across several edge nodes, the benefits of such an approach are exponential.

\begin{table*}[!ht]
\centering
\caption{A set of four operation modes used in our motivational pedestrian detection scenario.}
\label{tab:operation-modes}
\resizebox{\textwidth}{!}{%
\begin{tabular}{lllll}
\hline
\multicolumn{1}{c}{\multirow{2}{*}{\textbf{Operation Mode}}} &
  \multicolumn{1}{c}{\multirow{2}{*}{\textbf{Runtime Context}}} &
  \multicolumn{3}{c}{\textbf{Desirable Characteristics}} \\ \cline{3-5} 
\multicolumn{1}{c}{} &
  \multicolumn{1}{c}{} &
  \textit{\textbf{Energy Consumption}} &
  \textit{\textbf{Detection Accuracy}} &
  \textit{\textbf{Frames Processing Rate}} \\ \hline
\textit{\nopedestriansmode}  & no pedestrians detected             & very low & low      & moderate  \\ \hline
\textit{\fewpedestriansmode} & few pedestrians detected            & low      & moderate & moderate  \\ \hline
\textit{\smallgroupmode}     & small group of pedestrians detected & moderate & high     & high  \\ \hline
\textit{\crowdmode}           & crowd detected                      & high     & moderate & very high \\ \hline
\end{tabular}%
}
\end{table*}
 
\section{An Approach to Design Energy-Aware Self-Adaptive Applications}
\label{sec:approach}

\begin{figure*}[!th]
    \centering
    \includegraphics[width=0.9\textwidth]{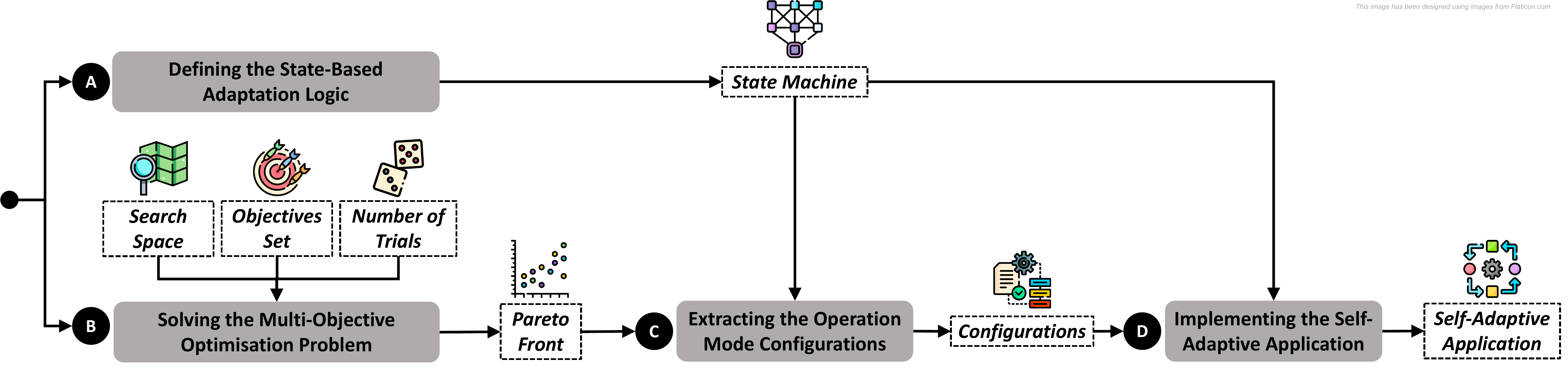}
    \caption{The steps of the proposed approach represented as a workflow diagram.}
    \label{fig:approach-workflow}
\end{figure*}

A \emph{self-adaptive application (SAA)} is an application capable of modifying itself or other connected resources in response to a continuously changing operational environment~\cite{brun2009engineering,salehie2012towards}. 

A SAA consists of a pair $(AL, MR)$, where $AL$ is the \emph{adaptation logic}, and $MR$ represents the \emph{managed resources}~\cite{krupitzer2015survey}, which are a group of resources, such as robotics, vehicles, and generic hardware with software, that the SAA can  control~\cite{krupitzer2015survey}. The adaptation logic is composed by all those items responsible for monitoring the environment (M), analyzing the data (A), planning (P), and executing the adaptation (E). This basic feedback framework proposed by Kephart and Chess~\cite{kephart2003vision} is named MAPE loop, and it is often extended by a knowledge component (K) responsible for managing content (e.g., monitoring values and adaptation policies).

SAAs are particularly effective in resource-constrained environments. We consider here the case of an AI-based application that implements the pedestrian detection use-case described in Section~\ref{sec:montivational-scenario} and that is hosted on  an embedded device (e.g., a Raspberry Pi) equipped with a video camera and a hardware accelerator (e.g., a TPU). The device executes an application capturing frames from the camera and processing them with an object detection model to detect pedestrians.

The hardware accelerator boosts the processing speed by lowering the ML model inference time. 
In this context, we must consider three main objectives: achieving high detection accuracy, processing frames at a high rate, and reducing energy consumption.

Optimizing these objectives at the same time for every possible operational condition is generally infeasible. Interestingly, a SAA can dynamically balance the degree of satisfaction of these objectives depending on the run-time context. However, engineers designing SAAs need to identify \emph{suitable configurations} for the run-time to balance the chosen objectives. Further, SAAs have to implement the logic to automatically switch between configurations (e.g., the four operation modes reported in Table~\ref{tab:operation-modes}), to adapt to changes in the operational environment (e.g., the pedestrian traffic volumes).

Identifying the configurations that implement the intended operation modes is also challenging, especially for AI-based applications running on heterogeneous and resource-constrained nodes.
Indeed, simply using a simulator may lead to results largely diverging from the real behavior of these applications. On the other hand, taking empirical measures by running the real devices and applications can be extremely expensive, especially when large configuration spaces must be explored~\cite{panerati2017optimization}.
We propose here an approach that combines the benefits of the \emph{empirical identification} of the configurations and those of an intelligent \emph{exploration of the configuration space} to yield suitable solutions to design an effective and energy-aware SAA.

Figure~\ref{fig:approach-workflow} describes our approach with a workflow diagram. An engineer provides the adaptation logic (A) as a finite-state machine (FSM) whose states represent the SAA operation modes and whose transitions encode the switching conditions between them.
In parallel, the engineer identifies the configuration space to explore, and defines a Multi-Objectives Optimization Problem (MOOP) that can be solved automatically (B) using a meta-heuristic search procedure.  Furthermore, the engineer specifies weights and thresholds for the objectives to guide the (C) extraction of the configurations to set in each operation mode.  The workflow terminates (D) with the implementation of the final FSM.

In the next subsections, we describe each step of the workflow in detail and exemplify the approach with the pedestrian detection scenario described in Section~\ref{sec:montivational-scenario}.

\subsection{Defining the  State-Based Adaptation Logic}
The first step of our approach requires an engineer, supported by domain experts, to define, in a rigorous way, the \emph{behavioral model} of the self-adaptive application~\cite{ghezzi2012runtime}.

As specification we use a \emph{Finite-State Machine (FSM)}, since it allows to explicitly represent the adaptation logic of an SAA~\cite{karsai1999model,arcaini2017formal,lee2019self}: the states represent the operational modes of the SAA, and the transitions represent the conditions triggering a change in the operation mode of the application. 

Formally, an FSM $M$ is defined by a tuple $(S, \Sigma, \delta, s_0)$, where $S$ is the set of states, $\Sigma$ is the set of the input symbols, that is, the set of events that may trigger state transitions, $\delta$ is the set of all the possible transitions from a state $s_1 \in S$ to a state $s_2 \in S$ caused by an event $\sigma \in \Sigma$, $s_0$ is the initial state.

\begin{figure}[!th]
    \centering
    \includegraphics[width=\columnwidth]{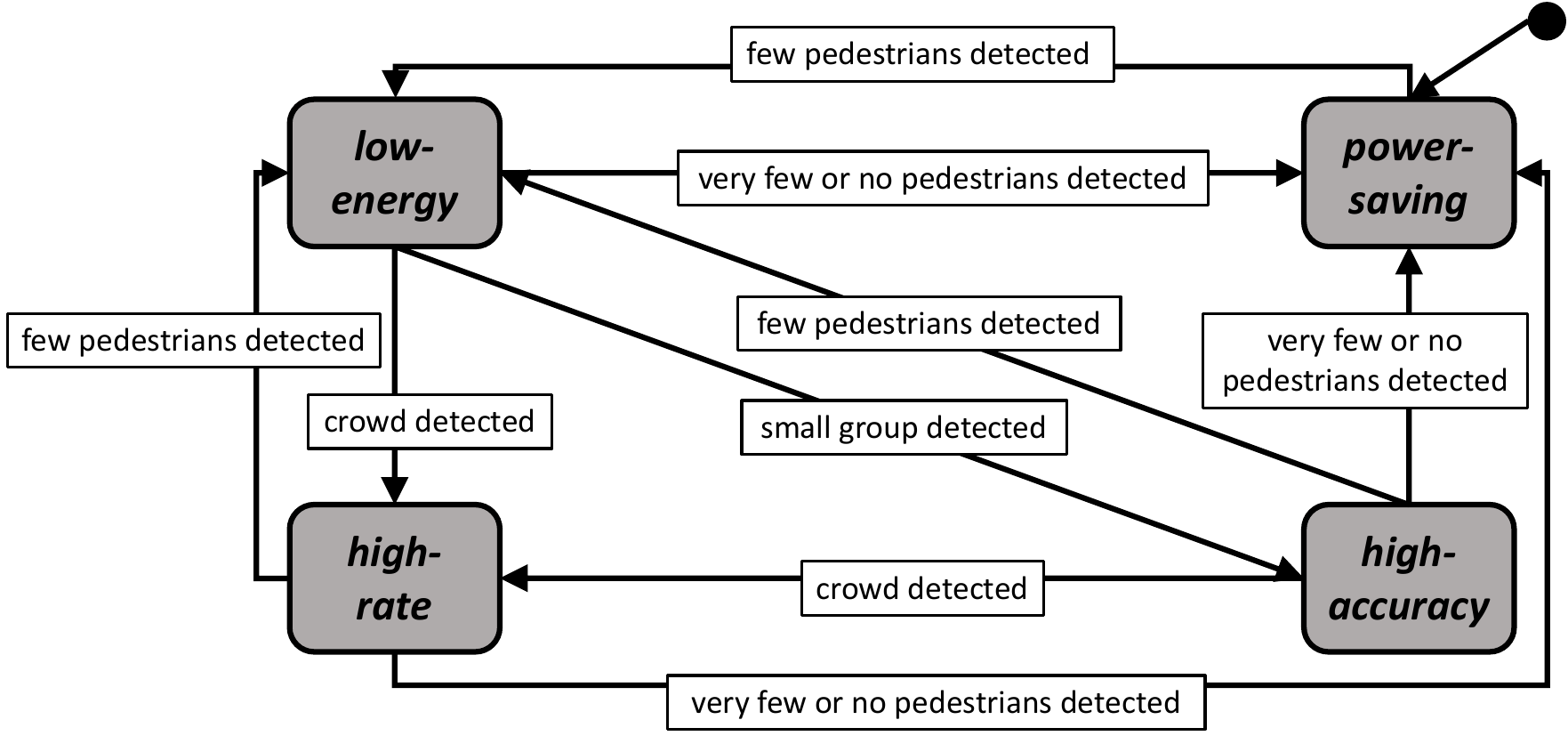}
    \caption{An abstract state machine modeling the states and the transitions of a self-adaptive application for our scenario.}
    \label{fig:abstract-fsm}
\end{figure}

Let us consider the pedestrian detection scenario again. Here an engineer may want to define a SAA that can self-adapt across four operation modes (see Table~\ref{tab:operation-modes}) to address the four possible run-time contexts in the area where the camera shall be deployed, defined for instance according to the available studies~\cite{aultman2009assessing,kim2017analysis,dobler2021patterns}.
Each operation mode, for example \emph{\fewpedestriansmode}, represents the working condition of the software that is best suited for the corresponding run-time context, for example \emph{few pedestrians detected}. Each operation mode must satisfy certain characteristics in terms of energy consumption, detection accuracy and frames processing rate. These characteristics are used to identify the exact software configurations at step (C) Extracting the Operation Mode Configurations \revised{MR.C2, R1.C3, R2.C3}{by providing the corresponding sets of objective weights and thresholds}.
Figure~\ref{fig:abstract-fsm} shows an abstract FSM, with the four identified abstract states and 9 transitions that capture when the software must self-adapt. 
\revised{R1.C5}{Please note that the domain-knowledge is exploited here to determine the transitions that must be encoded in the FSM, among the full set of the possible state transitions.}

\subsection{Solving the Multi-Objective Optimization Problem}
Finding high-quality software configurations that correspond to the operation modes identified by the engineer (e.g., the four states shows in Figure~\ref{fig:abstract-fsm}) is a hard problem. AI-based applications can be configured according to several parameters (see for instance the list of parameters that may influence pedestrian detection listed in Table~\ref{tab:parameters-domain}), generating a huge exploration space that cannot be exhaustively explored. Computer-simulated experiments can reduce the time and effort, but they are usually inaccurate, especially in Cyber Physical Systems and other domains that include real-world metrics~\cite{rupprecht2022survey}.
To address this challenge, we defined a \emph{Multi-Objective Optimization Problem (MOOP)} that is able to discover the configurations that deliver the best results for the considered set of objectives, and that can be exploited to 
find the actual configurations that effectively implement the operation modes represented as states of the FSM.

An optimization process aims to find a set of input values for a problem to obtain the ``\emph{optimal}'' output values. The definition of optimality is problem-specific, and formally, it refers to minimizing or maximizing one or more objective functions by varying the input values. Hence, a MOOP requires the satisfaction of a number of different and often conflicting objectives at the same time~\cite{stadler1988multicriteria,ngatchou2005pareto}. Intuitively, there is no single best solution for all the objectives, but rather there exist several optimal solutions representing the best trade-offs among all the objectives~\cite{stadler1988multicriteria}. The set of all possible solutions constitutes the \emph{search space}, which then also contains the set of input values revealing optimal outputs.

We define the search space $X$ as a set \emph{configurations}. A configuration $\mathit{conf}$ is $n$-tuple $(c_1, \dots, c_n)$, where $c_k$ is the value of the $k$-th configurable parameter $p_k \in P$ assuming values in its domain $D_{p_k}$. The size of X is $|X| = \prod_{k=1}^{n} |D_{p_k}|$.


The set of solutions $X^*$ is called the Pareto front, which contains all the solutions where no improvement is possible in any objective function without sacrificing at least one of the other objective functions~\cite{ngatchou2005pareto}. This is also referred to as the non-dominated solutions set. 

In the pedestrian detection scenario we have three objectives: (i) maximize the pedestrians detection accuracy (\emph{acc}), (ii) minimize the energy consumption (\emph{eng}), and (iii) maximize the number of processed frames in a time window (\emph{rate}). Hence, we define a MOOP with these three objectives (depending on the specific case, we might have a different number of objectives):  
\begin{mini}|s|
{}{-\mathit{acc}(\mathit{conf}) \wedge \mathit{eng}(\mathit{conf}) \wedge -\mathit{rate}(\mathit{conf})}{\label{eq:moop-scenario}}{}
\addConstraint{\mathit{conf} \in X}
\end{mini}
The search space $X$ is defined as a set of configuration quintuples with five configuration parameters for our application, that is, the camera resolution ($R$), the camera frame rate ($\mathit{FPS}$), the object detection model ($M$), the detection threshold ($T$), and whether to use the external hardware accelerator ($\mathit{TPU}$). Each parameter domain has a different cardinality (see details in Table~\ref{tab:parameters-domain}). Accordingly, $|X| = |R| \times |\mathit{FPS}| \times |M| \times |T| \times |\mathit{TPU}|=3402$ configuration quintuples.

\begin{table*}[!th]
\centering
\caption{The domain of the parameters used to define the search space of the multi-objective optimization problem.}
\label{tab:parameters-domain}
\resizebox{\textwidth}{!}{%
\begin{tabular}{@{}lll@{}}
\toprule
\textbf{Parameter}                  & \textbf{Parameter Type}                    & \textbf{Domain}                                 \\ \midrule
Camera Resolution ($R$)             & Categorical                                & \{1920x1080, 1280x720, 640x480\}                \\ \midrule
Camera Frame Rate ($\mathit{FPS}$)  & Categorical                                & \{1, 5, 10, 15, 20, 25, 30\}                    \\ \midrule
Object Detection Model ($M$) &
  Categorical &
  \begin{tabular}[c]{@{}l@{}}\{SSD MobileNet V1, SSD/FPN MobileNet V1 TF2, SSD MobileNet V2,\\ SSD MobileNet V2 TF2, SSDLite MobileDet, EfficientDet-Lite0,\\ EfficientDet-Lite1, EfficientDet-Lite2, EfficientDet-Lite3\}\end{tabular} \\ \midrule
Detection Threshold ($T$)           & Numerical (low: 0.1, high: 0.9, step: 0.1) & \{0.1, 0.2, 0.3, 0.4, 0.5, 0.6, 0.7, 0.8, 0.9\} \\ \midrule
Use HW Accelerator ($\mathit{TPU}$) & Categorical                                & \{true, false\}                                 \\ \bottomrule
\end{tabular}%
}
\end{table*}

Solving the Eq.~\ref{eq:moop-scenario} results in a Pareto front with non-dominated solutions, that is, configurations that \emph{fulfill the three objectives by a different, but relevant, degree}. We use a strategy derived from NSGA-II to compute the Pareto front.



NSGA-II is a solid, fast, and widely used optimization algorithm in real-world applications~\cite{verma2021comprehensive}. We use the approach defined by Deb et al.~\cite{deb2002fast} for the exploration of the search space: it is explored by searching for dominant solutions (i.e., the fitness of a solution is defined by computing its non-domination level) in less populated areas of the space (i.e., determined by computing the crowding distance) guaranteeing the diversity of the identified solutions; mutations randomly change parameter values with a probability that is computed according to the number of parameters in the configuration, and uniform crossover recombines configurations with a probability of 0.9.

During the search space exploration, our procedure records all the evaluated objective values, and at the end it extracts the Pareto front from the whole results set. In the empirical evaluation, we show how this strategy can be used to explore only 10\% of the search space to select nearly optimal configurations. Note this is particularly relevant, since assessing how a single configuration fulfills the three objectives requires collecting empirical measures by repeating a same experiment multiple times. 



\subsection{Extracting the Operation Mode Configurations}

The Pareto front obtained by solving the MOOP usually contains a large number of non-dominated solutions, compared to the operation modes needed by the self-adaptive application. 
The decision-making process to identify the actual solutions from the Pareto front involves comparing multiple criteria, trading-off certain objectives for others~\cite{kuo2008use, wang2017application}. 
To address this problem, we use  the \emph{weighted gray relational analysis (WGRA)}~\cite{kuo2008use} method, a weighted version of the GRA introduced by Ju-Long~\cite{ju1982control} and employed in multiple application domains~\cite{chia2021weighted}. 
This is a very robust method~\cite{mahmoudi2020distinguishing}, preferable to other multi-criteria decision making (MCDM) methods as it inherently incorporates uncertainty in data, and it is simple to calculate~\cite{mahmoudi2020distinguishing, wu2002comparative} and to integrate into existing software.

GRA combines into a single value all the objectives. This simplifies the original MCDM problem into a single-criterion decision-making problem~\cite{kuo2008use}, making Pareto front solutions easily comparable.
To let engineers extract states that fulfill the objectives by different degrees, we employ the weighted version of the algorithm that uses a set of weights $W$ to give more importance to certain objectives~\cite{chia2021weighted}.

The WGRA algorithm consists of three main steps: (i) data normalization, (ii) reference network computation, and (iii) gray relational grade (GRG) computation~\cite{wang2017application}.

The \emph{data normalization} step consists of the normalization of the objective values in the Pareto front according to two cases: larger-the-better for maximization, and smaller-the-better for minimization.  The normalized value $F_{ij}$ is calculated by Eq.~\ref{eq:normalize-max-objective} and~\ref{eq:normalize-min-objective} for maximization and minimization cases, respectively:
\begin{align}
& F_{ij} = \frac{f_{ij} - \text{min}_{i \in n}f_{ij}}{\text{max}_{i \in n}f_{ij} - \text{min}_{i \in n}f_{ij}}\label{eq:normalize-max-objective} \\
& F_{ij} = \frac{\text{max}_{i \in n}f_{ij} - f_{ij}}{\text{max}_{i \in n}f_{ij} - \text{min}_{i \in n}f_{ij}}\label{eq:normalize-min-objective}
\end{align}
with $f_{ij}$ as the $i$-th value of the $j$-th objective in the matrix $O$, a matrix $n \times m$ composed of $n$ Pareto front solutions and $m$ objectives. $F_{ij}$ is the value of $f_{ij}$ after normalization.

The \emph{reference network computation} step consists in forming the reference network $F^{+}_{j}$, that is, an ideal network obtained by choosing the best value of each of the objectives as follows:
\begin{equation}
    F^{+}_{j} = \text{max}_{i \in n}F_{ij}
\end{equation}

Finally, the \emph{gray relational grade (GRG) computation} step consists in calculating the similarity between each candidate network (i.e., the objective values of each optimal solution in the Pareto front) and the reference network $F^{+}_{j}$. The GRG for each $i$-th value in the Pareto Front is computed as follows:
\begin{equation}\label{eq:grc}
    GRG_{i} = \frac{1}{n} \sum_{j=1}^{m} w_{j}\frac{\Delta\text{min} - \Delta\text{max}}{\Delta_{ij} + \Delta\text{max}}
\end{equation}
where $w_j$ is the weight of the $j$-th objective value (with $\sum_{j=1}^{m} w_{j} = 1$); $\Delta{ij} = |F^{+}_{j} - F_{ij}|$ is the absolute value of the difference of between the $j$-th objective value in the reference network and the one in the candidate network; $\Delta\text{max} = \text{max}_{i \in n, j \in m}(\Delta{ij})$ and $\Delta\text{min} = \text{min}_{i \in n, j \in m}(\Delta{ij})$ are the maximum and minimum deltas, respectively.

The $\mathit{conf} \in X$ with the largest $\text{GRG}_{i}$ is the recommended optimal solution outputed by the WGRA process. Depending on the set of weights used to extract the configuration from the Pareto front, the configuration shall map to a different state of the FSM, that is, it implements a different operation mode of the AI-based edge service.



To illustrate further, let us focus on two operation modes in our example, namely, \emph{\nopedestriansmode} and \emph{\crowdmode}. The engineer, jointly with domain experts~\cite{hsu2011comparison}, may provide the following sets of weights for the two operation modes, respectively: $W_{\textit{\nopedestriansmode}} = \{0.05, 0.9, 0.05\}$ and $W_{\textit{\crowdmode}} = \{0.6, 0, 0.4\}$. 
The specific weights could be derived from a Service Level Agreement (SLA) defining the QoS, and the costs the application service provider to sustain and deliver the application.

Engineers could also define a set of objective thresholds $t_{j}$ for each objectives $O_j$  to reduce the size of the Pareto front given in input to the WGRA algorithm, filtering out solutions that might be unreasonable for a given operation mode $op$. In particular, a solution is filtered from the Pareto front if the value it achieved on objective $O_j$ is above the threshold $t_{j}$. 

For example, let us consider the \emph{\nopedestriansmode} and the \emph{\crowdmode} operation modes again. The weights assigned to the $W_{\textit{\nopedestriansmode}}$ set must give a large importance to the energy consumption objective in order to extract an energy-efficient configuration. However, this may lead to the identification of a very poor but still non-dominated solution for the other two objectives. To prevent this risk, the engineer can filter all the solutions that do not provide a minimum detection accuracy level and/or number of processed frames. For instance, they can define a set of thresholds $T_{\textit{\nopedestriansmode}} = \{ t_{acc}, t_{eng}, t_{rate} \} = \{ 0.2, 0, 60 \}$ to exclude solutions with a detection accuracy lower than 0.2, and a number of processed frames lower than 60. A completely different set of thresholds could be defined for the \emph{\crowdmode}, that is, $T_{\textit{\crowdmode}} = \{ 0.3, 0, 0 \}$. In this case, solutions with a detection accuracy lower than 0.3 are filtered out in order to provide a minimum detection accuracy level, when compared to the \emph{\nopedestriansmode} mode.


\begin{figure}[!th]
    \centering
    \includegraphics[width=\columnwidth]{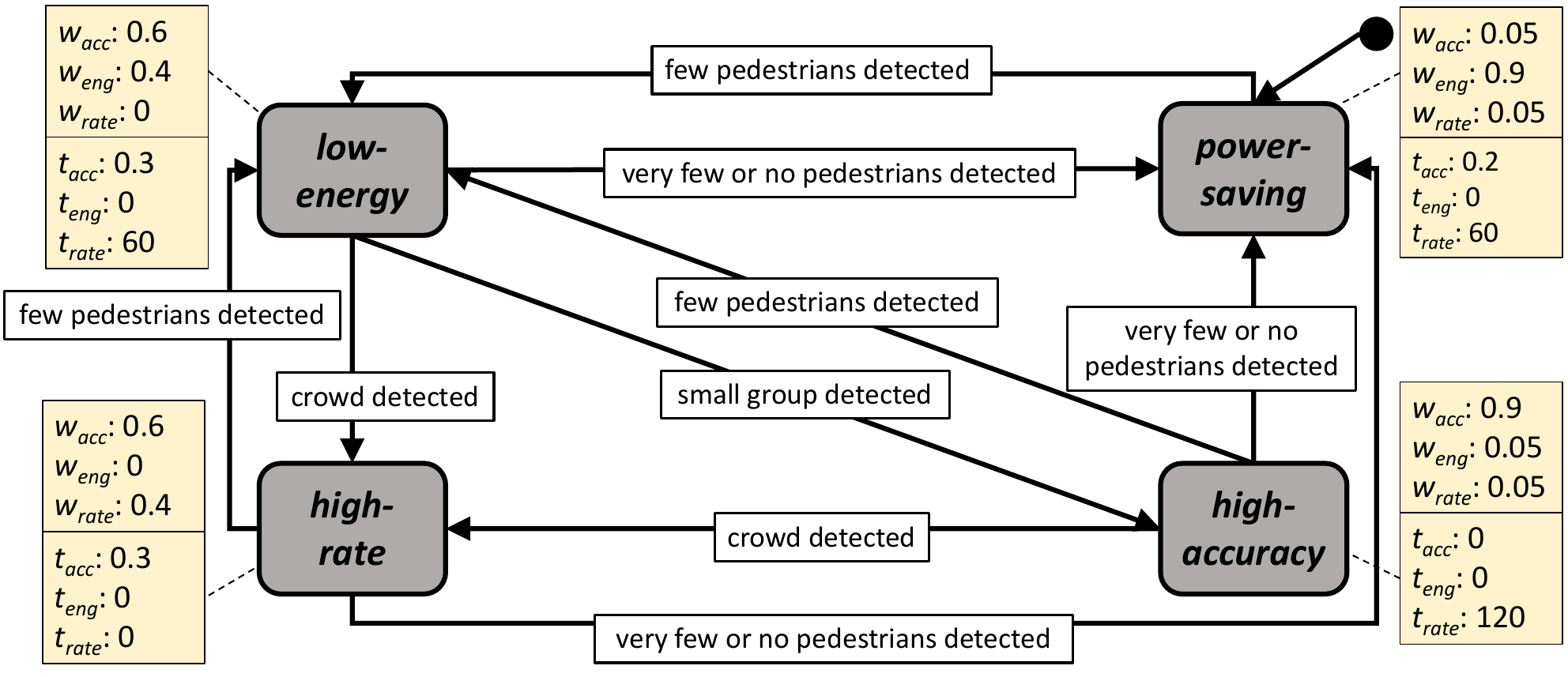}
    \caption{A refined version of the abstract state machine shown in Figure~\ref{fig:abstract-fsm} with the set of weights and thresholds for each of the operation modes.}
    \label{fig:refined-fsm}
\end{figure}


Figure~\ref{fig:refined-fsm} shows the refined version of the abstract FSM previously shown in Figure~\ref{fig:abstract-fsm} with the weights and thresholds for WGRA analysis defined by the engineers attached to states. \revised{MR.C2, R1.C3, R2.C3}{The chosen weights and thresholds represent the actual specification of the  \emph{desirable characteristics} of the operation modes listed in Table~\ref{tab:operation-modes}.} The execution of the WGRA algorithm for each of the FSM state extracts a configuration $\mathit{conf}_{op}$ with the actual configuration parameter values that can be used by the SAA application to self-adapt the operation mode.

\subsection{Implementing the Self-Adaptive Application}
In the last step, the engineer is required to  implement the self-adaptive application according to the output of the analysis. The abstract state machine is transformed into a concrete one in two steps: first, each of the transitions must be turned into an actual triggering condition; second, the operation mode configurations extracted in the previous step are mapped into a piece of logic able to set these configurations at runtime.
Fig.~\ref{fig:concrete-fsm} shows the final FSM for our pedestrian detection scenario, with actual conditions and operation modes. 

\begin{figure}[!th]
    \centering
    \includegraphics[width=\columnwidth]{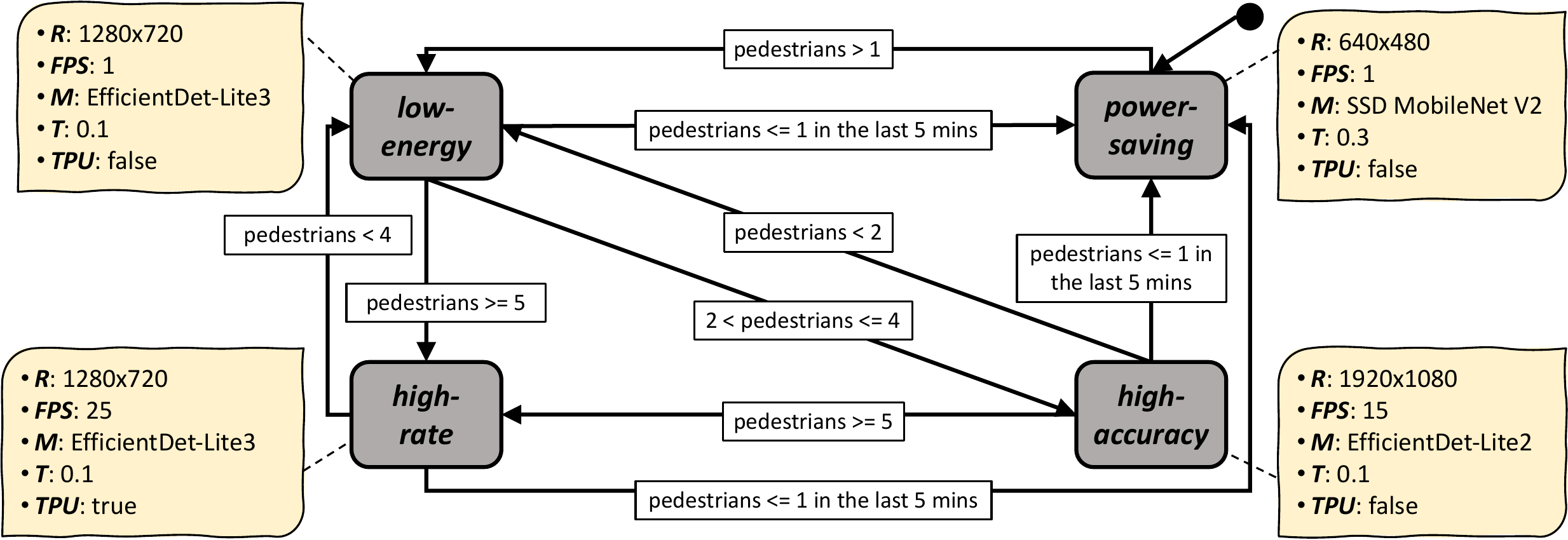}
    \caption{The concrete finite state machine implementing a self-adaptive application for our scenario.}
    \label{fig:concrete-fsm}
\end{figure}

The FSM can be translated into working code using generators~\cite{pruteanu2008genfsm, vandeportaele2017finite} or when this is not possible or too difficult~\cite{adamczyk2004selected}, the SAA can be obtained semi-automatically or manually~\cite{adamczyk2003anthology,adamczyk2004selected,yacoub2004pattern}. Our approach outputs a concrete FSM encoding the SAA and does not bind the engineer to use any specific method to implement the SAA.

 
\section{Empirical Evaluation}
\label{sec:evaluation}
To evaluate our approach, we investigate the following two research questions in the context of the pedestrian detection case study described in Section~\ref{sec:montivational-scenario}. We select such a case study since it represents a real-world and challenging scenario that requires delivering effective and sustainable AI edge services. 

\noindent \textbf{RQ1 (Meta-Heuristic VS Near-Exhaustive Search) - Can our meta-heuristic search approach discover solutions whose quality is comparable to those obtained by a near-exhaustive search?} This research question investigates the effectiveness of our meta-heuristic strategy. In particular, it studies whether the heuristic exploration of a small portion of the search space can lead to results comparable to a near-exhaustive exploration. 

\noindent \textbf{RQ2 (Objectives Trade-Off) - Can a self-adaptive pedestrians detection application better balance energy consumption and application objectives compared to a non-adaptive application?} This research question investigates whether the self-adaptive application resulting from our methodology can release a better trade-off among accuracy, energy, and processing speed compared to \revised{R2.C4}{four baseline non-adaptive applications}. 

\subsection{Experimental Setting}
\label{subsec:experimental-setting}

\begin{figure}[!th]
     \centering
     \includegraphics[width=0.9\columnwidth]{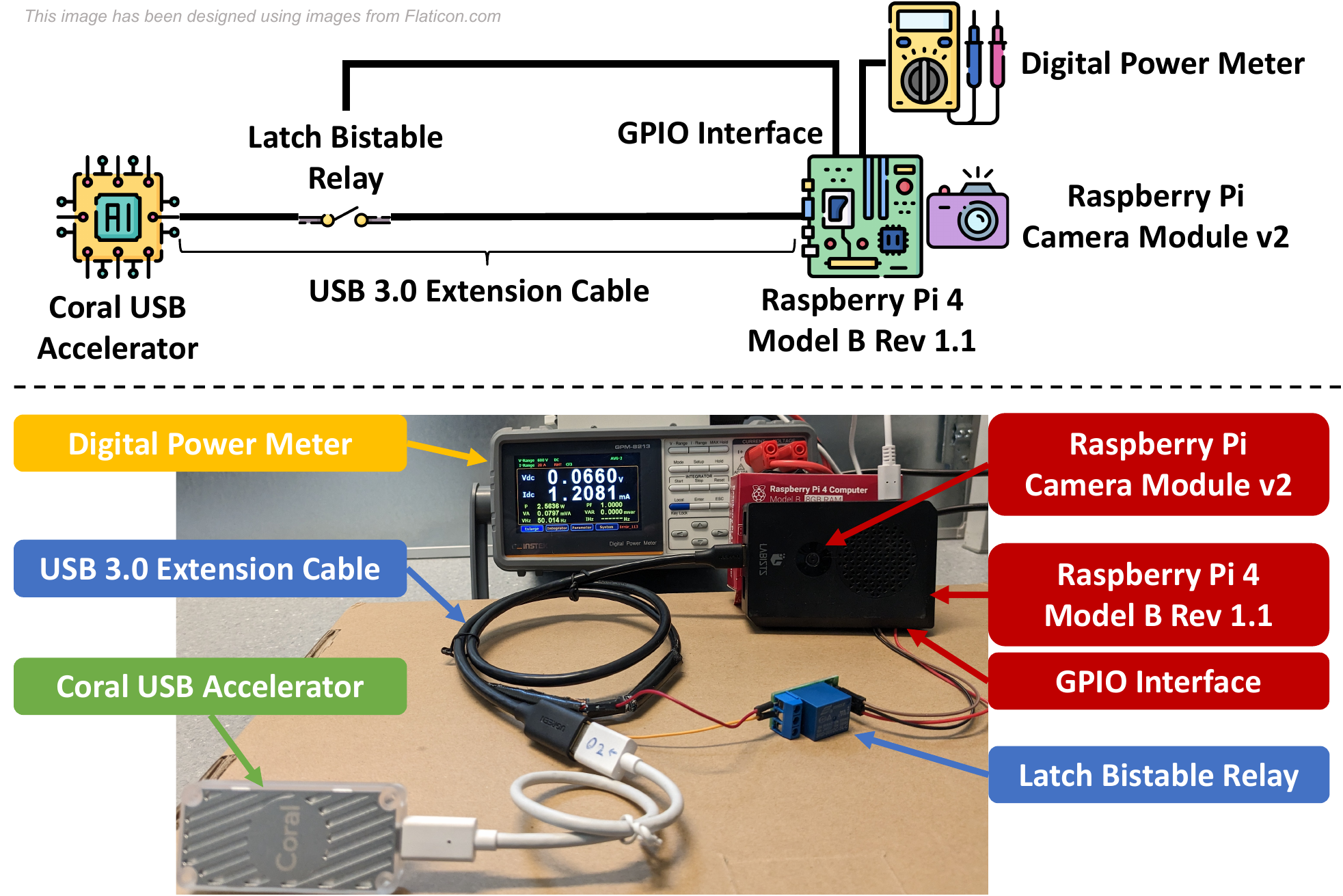}
     \caption{The test-bed used to run the evaluation experiments.}
     \label{fig:testbed}
\end{figure}

Fig.~\ref{fig:testbed} shows the test-bed we used to run our case study evaluation, first schematically (above), then its concrete in-lab implementation (below).

We employ a Raspberry Pi (RPi) 4 Model B Rev 1.1 (64-bit quad-core ARMv8, 4GB of RAM, RPi OS Lite 64-bit Debian GNU/Linux 11) equipped with the RPi Camera Module v2 and boxed in a LABISTS case\footnote{\url{https://labists.com/products/raspberry-pi-4-case-kit}} with a 5V fan connected to the RPi General Purpose Input/Output (GPIO) interface. The RPi is powered by a USB-C AC adapter connected through a GW Instek GPM-8213 digital power meter\footnote{\url{https://www.gwinstek.com/en-GB/products/detail/GPM-8213}} that we use to collect instant power values.

To reduce the idle energy consumption of RPi, we disable the unnecessary components: all the LEDs (i.e., activity, power, and Ethernet port), the Wi-Fi antenna, the Bluetooth, and the HDMI port. Internet and private network connectivity is provided via network cable.
A Coral USB Accelerator (Edge TPU)\footnote{\url{https://coral.ai/products/accelerator}} is plugged-in for those experiments that require hardware accelerator. The accelerator is automatically powered-on when connected to the USB port. 



Since there is no possibility to enable and disable a single USB port on-the-fly via software, a self-adaptive application running on such device would not be capable to completely power-off the accelerator when not in use, reducing the potential benefits of switching to an energy-efficient operation mode.  To overcome this limitation, we realize a software-level power switch by employing a latch bi-stable relay (SONGLE SRD-05VC-SL-C) connected to the GPIO interface and a USB 3.0 extension cable. 
This enables us to turn it on and off by triggering the relay through software to close or open the circuit using a GPIO pin.
For pedestrian detection, we employ state-of-art object detection models pre-trained on the COCO dataset~\cite{lin2014microsoft}. The models are publicly available at the Coral.ai website\footnote{\url{https://coral.ai/models/object-detection/}}, and they are already compiled for both CPU and Edge TPU execution. We reported detailed information to re-create our test-bed on our public repository \repo.



\subsection{RQ1 - Meta-Heuristic VS Near-Exhaustive Search}
\begin{figure*}[!th]
     \centering
     \begin{subfigure}[t]{0.245\textwidth}
         \centering
         \includegraphics[width=\textwidth]{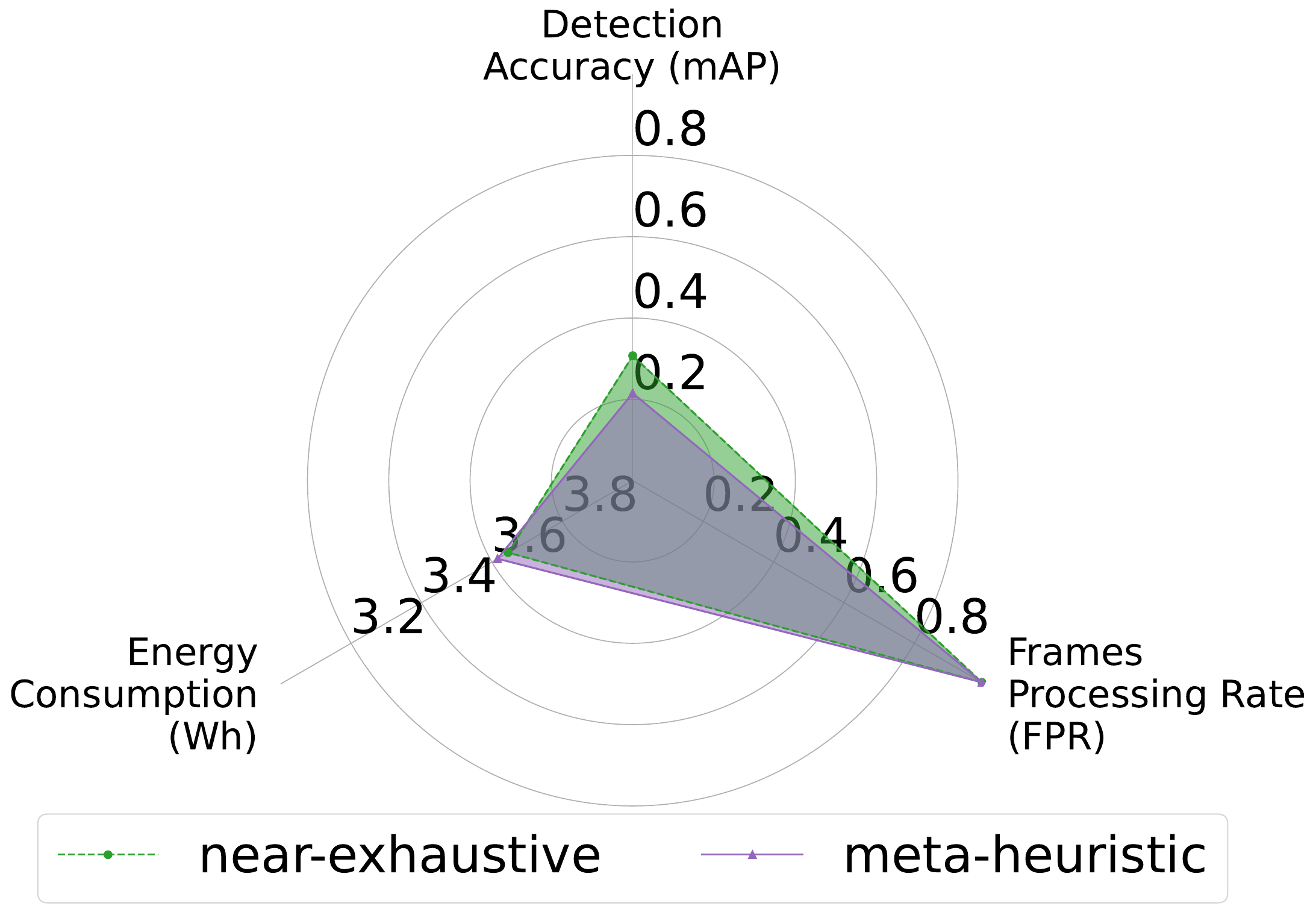}
         \caption{\nopedestriansmode}
         \label{fig:power-saving-radar-heuristic-vs-exhaustive}
     \end{subfigure}
     \hfill
     \begin{subfigure}[t]{0.245\textwidth}
         \centering
         \includegraphics[width=\textwidth]{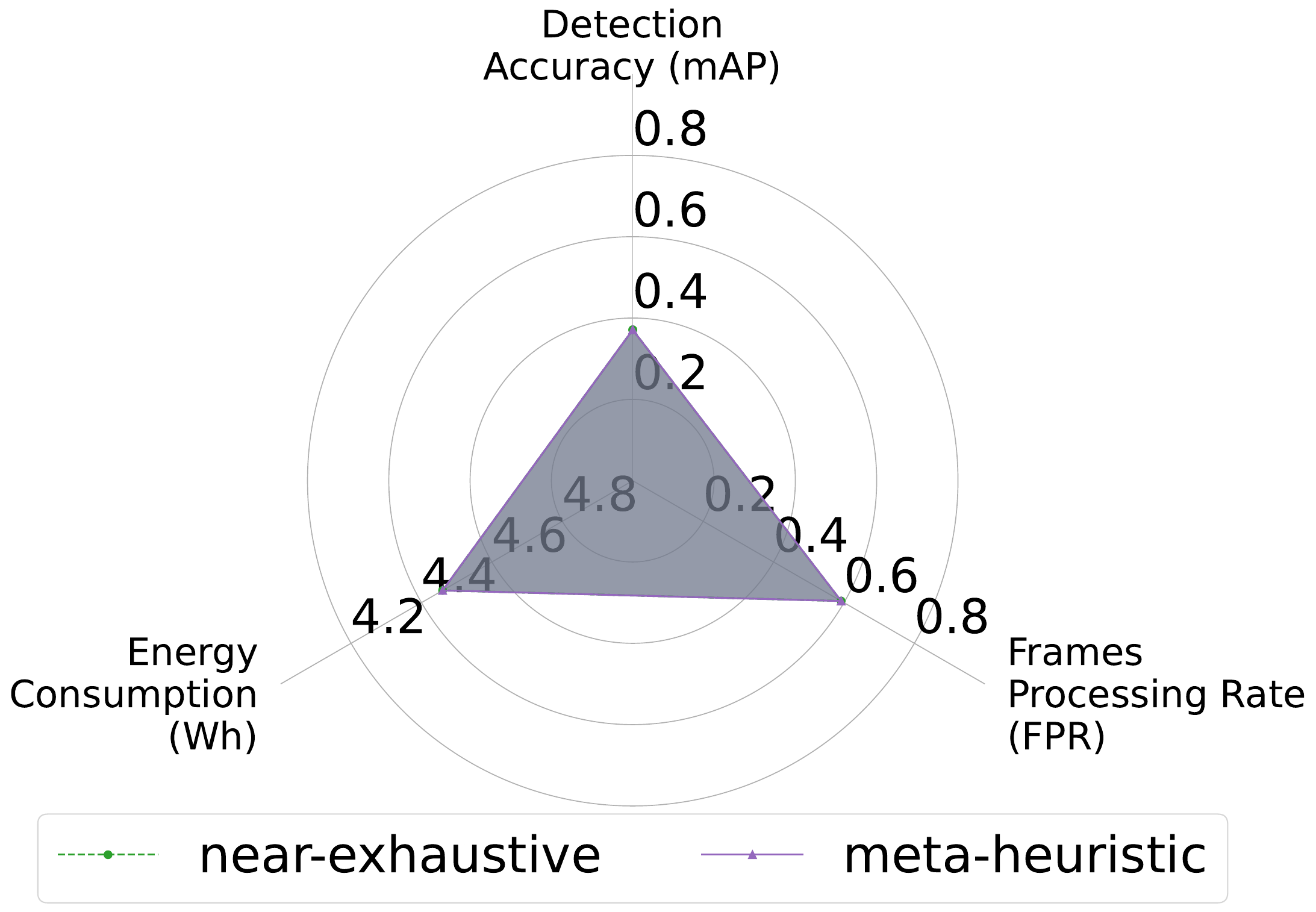}
         \caption{\fewpedestriansmode}
         \label{fig:low-energy-radar-heuristic-vs-exhaustive}
     \end{subfigure}
     \hfill
     \begin{subfigure}[t]{0.245\textwidth}
         \centering
         \includegraphics[width=\textwidth]{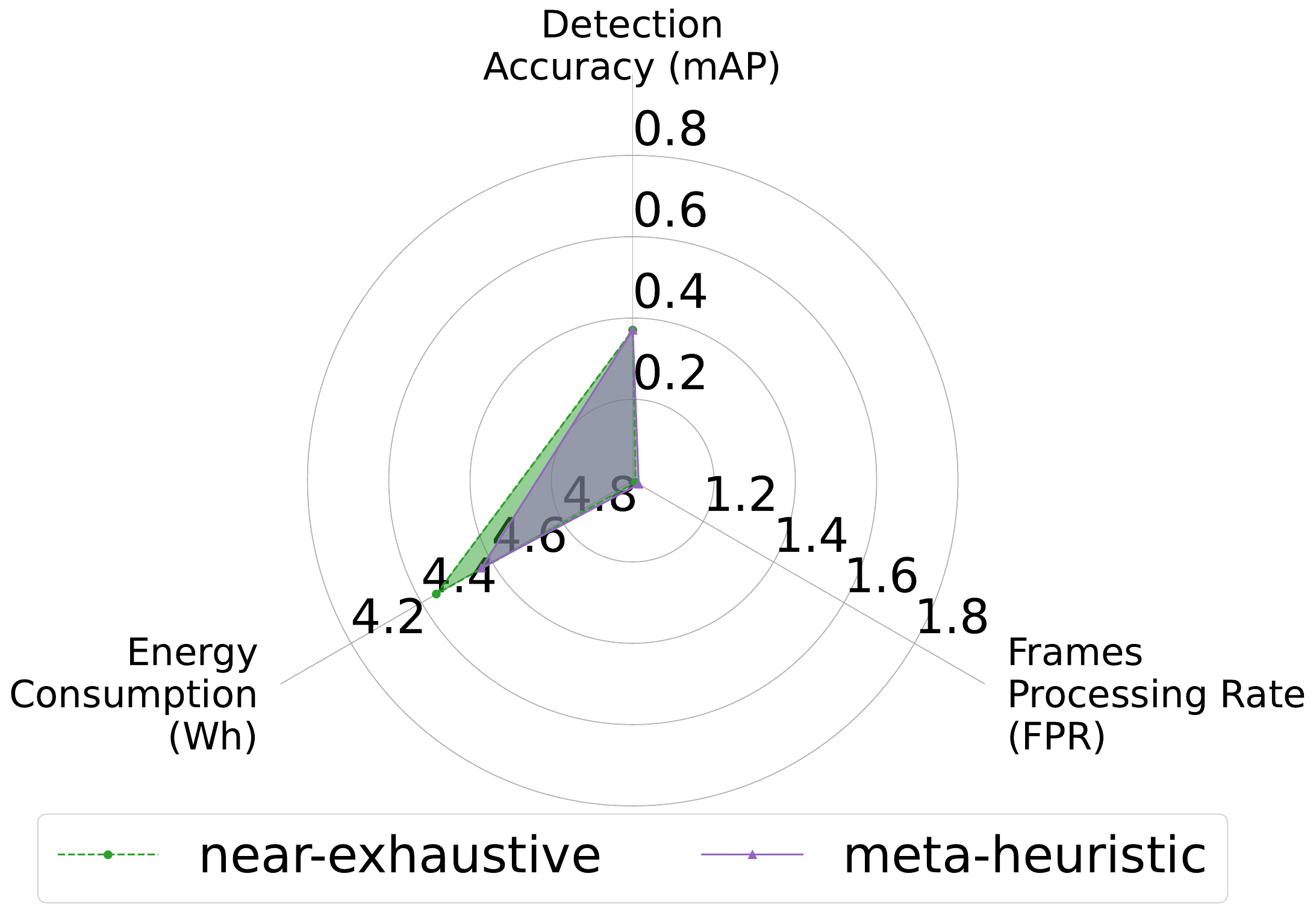}
         \caption{\smallgroupmode}
         \label{fig:high-accuracy-radar-heuristic-vs-exhaustive}
     \end{subfigure}
     \hfill
     \begin{subfigure}[t]{0.245\textwidth}
         \centering
         \includegraphics[width=\textwidth]{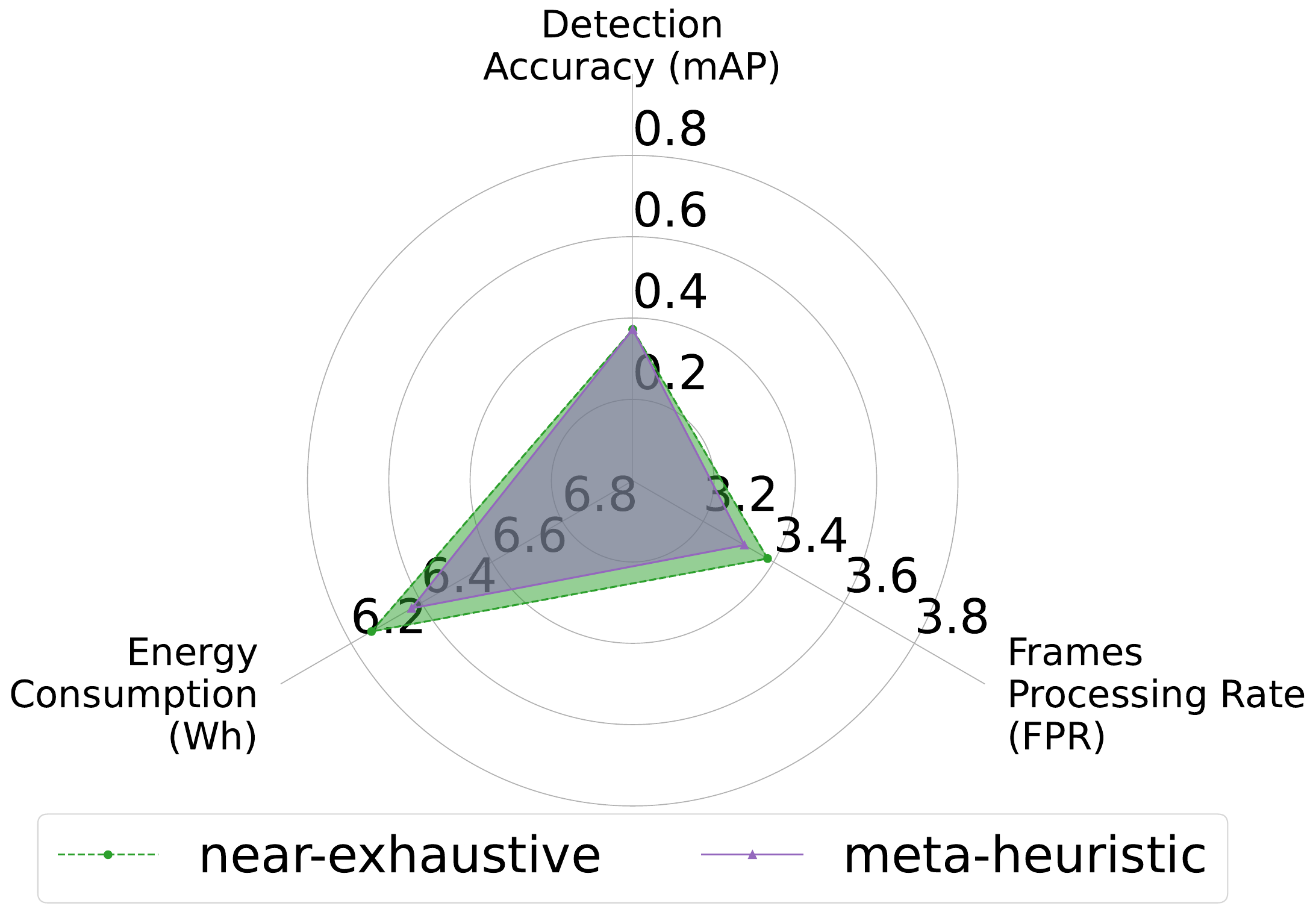}
         \caption{\crowdmode}
         \label{fig:high-rate-radar-heuristic-vs-exhaustive}
     \end{subfigure}
     \caption{Radar charts comparing the objective values of the four self-adaptive operation modes when employing a solution obtained with the meta-heuristic search procedure and one obtained with the near-exhaustive search procedure. The solutions are extracted with the WGRA method using the same set of weights and thresholds.}
     \label{fig:radars-heuristic-vs-exhaustive}
\end{figure*}

This research question aims to investigate whether exploring a small portion of the search space efficiently can lead to comparable results with a near-exhaustive exploration.

To answer RQ1, we first compute the Pareto front of the MOOP as defined in Eq.~\ref{eq:moop-scenario} with our meta-heuristic search procedure by only exploring 10\% of the search space reported in Table~\ref{tab:parameters-domain} (i.e., 340 unique trials out of 3402 trials), and then we explore more than 80\% of the same space (i.e., 2790 unique trials out of 3402 trials) with a random search procedure. Second, we extract the four operation modes needed to address the pedestrian detection scenario (according to the weights and thresholds reported in Fig.~\ref{fig:refined-fsm}) from the two Pareto fronts: the one computed with the meta-heuristic search procedure and the one obtained with the near-exhaustive procedure. Finally, we compare the objective values achieved with the two SAAs that derive from the two sets of selected states. A good meta-heuristic procedure should be able to achieve results as good as the near-exhaustive exploration.

The whole optimization procedure is implemented with the Optuna framework~\cite{optuna2019}, a state-of-art hyperparameter optimization framework with MOOP capabilities. Our meta-heuristic search procedure with memory capabilities is realized by using the NSGAIISampler\footnote{\url{https://optuna.readthedocs.io/en/stable/reference/samplers/generated/optuna.samplers.NSGAIISampler.html}} implementing the NSGA-II algorithm and the results database provided by Optuna. We use the default framework values to configure the sampler 
and we repeat the search 10 times with a different seed value recorded for reproducibility. The near-exhaustive search procedure, instead, employs the RandomSampler\footnote{\url{https://optuna.readthedocs.io/en/stable/reference/samplers/generated/optuna.samplers.RandomSampler.html}}.

At each optimization round, when a sampler selects a point $conf$ from the search space, two experiments must be executed to determine the objective values for the selected $conf$.

The first experiment computes the \emph{detection accuracy} by employing a pedestrian street scene belonging to the Multiple Object Tracking benchmark dataset~\cite{dendorfer2021motchallenge} (i.e., the ADL-Rundle-6 video). Both the frame size and the ground truth have been properly adjusted to match the camera resolution (R) parameter values defined by the search space. 
We use the Mean Average Precision (mAP) as detection accuracy metric, a popular metric for object detection algorithms~\cite{padilla2020survey}, and we evaluate the model predictions by using the open-source FiftyOne COCO-style evaluator\footnote{\url{https://docs.voxel51.com/user_guide/evaluation.html}}.

The second experiment, instead, computes both the achieved \emph{Frames Processing Rate} (FPR) and the \emph{energy consumption}. We run the pedestrian detection application on the device (i.e., the Raspberry Pi described in Section~\ref{subsec:experimental-setting} for 120 seconds configured according to \emph{conf}). We collect both 
the consumed energy in Watt-hours (Wh) 
and the FPR computed as the ratio between the number of processed frames and the experiment duration.

\paragraph*{Results}

The near-exhaustive search executed for about 18 days sampling 2790 unique trials and discovered a Pareto front with 131 solutions. The meta-heuristic search executed for about 54 hours sampling 340 unique trials (10\% of the entire space) and discovered a Pareto front with 83 solutions on average. Note that the saving, when the sampling involves running experiments, is significant in both relative and absolute terms (more than 2 weeks of computing saved).

Since each run of the meta-heuristic search may return a slightly different configuration for a given state, we selected the configuration that occurred most frequently in the 10 repetitions to derive the corresponding SAA. When multiple solutions have the same highest frequency, we excluded the solution matching the one extracted from the near-exhaustive Pareto front to avoid any bias, and consider a worst case scenario.

Fig.~\ref{fig:radars-heuristic-vs-exhaustive} shows four radar charts - one per each operation mode in the SAA - comparing the three objective values of the solution extracted with the near-exhaustive search (green, dashed, dot mark), and the one extracted from meta-heuristic search (purple, solid line, triangle mark), respectively. Each of the axes has its own scale, but for all the objectives, the higher is the value the better it is.

The plots clearly indicate that the states identified by our meta-heuristic search procedure and the ones obtained with the near-exhaustive search result in highly similar performance.
The \emph{\fewpedestriansmode} operation mode (Fig.~\ref{fig:low-energy-radar-heuristic-vs-exhaustive}) resulted in exactly the same solution returned by the two procedures. While the near-exhaustive search identified solutions performing comparably to the ones identified by the meta-heuristic search in the remaining three operation modes.

In the case of the \emph{\nopedestriansmode} operation mode (Fig.~\ref{fig:power-saving-radar-heuristic-vs-exhaustive}), the two solutions perform with the same FPR and with negligible difference in energy consumption ($< 1\%$). The difference is slightly larger for the detection accuracy (0.307 mAP VS 0.215 mAP), whose relevance in the power-saving mode is however limited.

In the case of the \emph{\smallgroupmode} operation mode (Fig.~\ref{fig:high-accuracy-radar-heuristic-vs-exhaustive}), the two solutions perform with the same detection accuracy, and with negligible differences for FPR($< 1\%$) and energy consumption (4.442 Wh VS 4.570 Wh).

Finally, the two solutions obtained for \emph{\crowdmode} (Fig.~\ref{fig:high-rate-radar-heuristic-vs-exhaustive}) perform with the same detection accuracy, and with negligible differences for both FPR and the energy consumption ($< 2\%$). 

We can conclude that our search procedure has been as effective as the near exhaustive procedure for the pedestrian detection scenario, despite an empirical exploration of only 10\% of the search space.

\subsection{RQ2 - Objectives Trade-Off}
This research question aims to investigate whether a self-adaptive application changing its operation mode can better balance  the fulfillment of multiple objectives compared to a non-adaptive application using a single operation mode.

We study this research question in the context of two pedestrian traffic scenarios, namely, \emph{weekdays} and \emph{weekends}, derived from real-world traffic shapes reported by Dobler et al.~\cite{dobler2021patterns} in their work about urban pedestrians dynamic in the borough of Manhattan. In particular, the \emph{weekdays} scenario has a 3-peaks structure aligned with the ``9-to-5'' workday time, in which the peaks correspond to commuting to work, exiting buildings at lunch time, and leaving the work place. The \emph{weekend} scenario does not show a peaked structure, but rather a steady increase of pedestrians until the night.

We create a scenario by selecting 1440 frames, that is, 60 frames per hour, from a pool 115 of manually annotated frames containing between 0 and 5 pedestrians. Each hour of the day is labeled as 0 pedestrians, 1 to 3 pedestrians, and 4 to 5 pedestrians. The frames used for the experiment are taken from a study about real-time analytics for traffic safety~\cite{lujic2021increasing}. 

We implement a self-adaptive pedestrian detection application according to the FSM depicted in Fig.~\ref{fig:concrete-fsm} using the Python State Machine library ({\small \url{https://pysm.readthedocs.io/}}). Then, we use the same pedestrian detection logic to obtain the non-adaptive baseline application. The four operation mode configurations obtained by the meta-heuristic search procedure in RQ1 are used to configure both the self-adaptive 
\revised{R2.C4}{application and the non-adaptive baselines, obtaining four non-adaptive applications.}
Fig.~\ref{fig:concrete-fsm} shows the configuration parameter values. Further, we include in the study a non-adaptive configuration, namely the \emph{balanced} configuration, that assigns the same weight ($0.33$) to the three objectives and uses the thresholds ($t_{acc}=0.3, t_{eng}=0, t_{rate}=120$) that filter out the same unsatisfactory configurations collectively filtered out by the four operation modes of the adaptive approach. This configuration implements the best attempt to balance all the objectives without introducing any self-adaptation logic. Interestingly, the \emph{balanced} configuration matches our \emph{high-accuracy} configuration, that is, high-accuracy can be released maintaining a good level of energy consumption and frame rate.  

We evaluate the resulting self-adaptive and non-adaptive applications by using the same set of metrics used for RQ1, that is, the MOOP objectives: detection accuracy (mAP), energy consumption (Wh), and FPR. 

\paragraph*{Results}
\begin{figure}[!th]
     \centering
     \begin{subfigure}[t]{0.8\columnwidth}
         \centering
         \includegraphics[width=\textwidth]{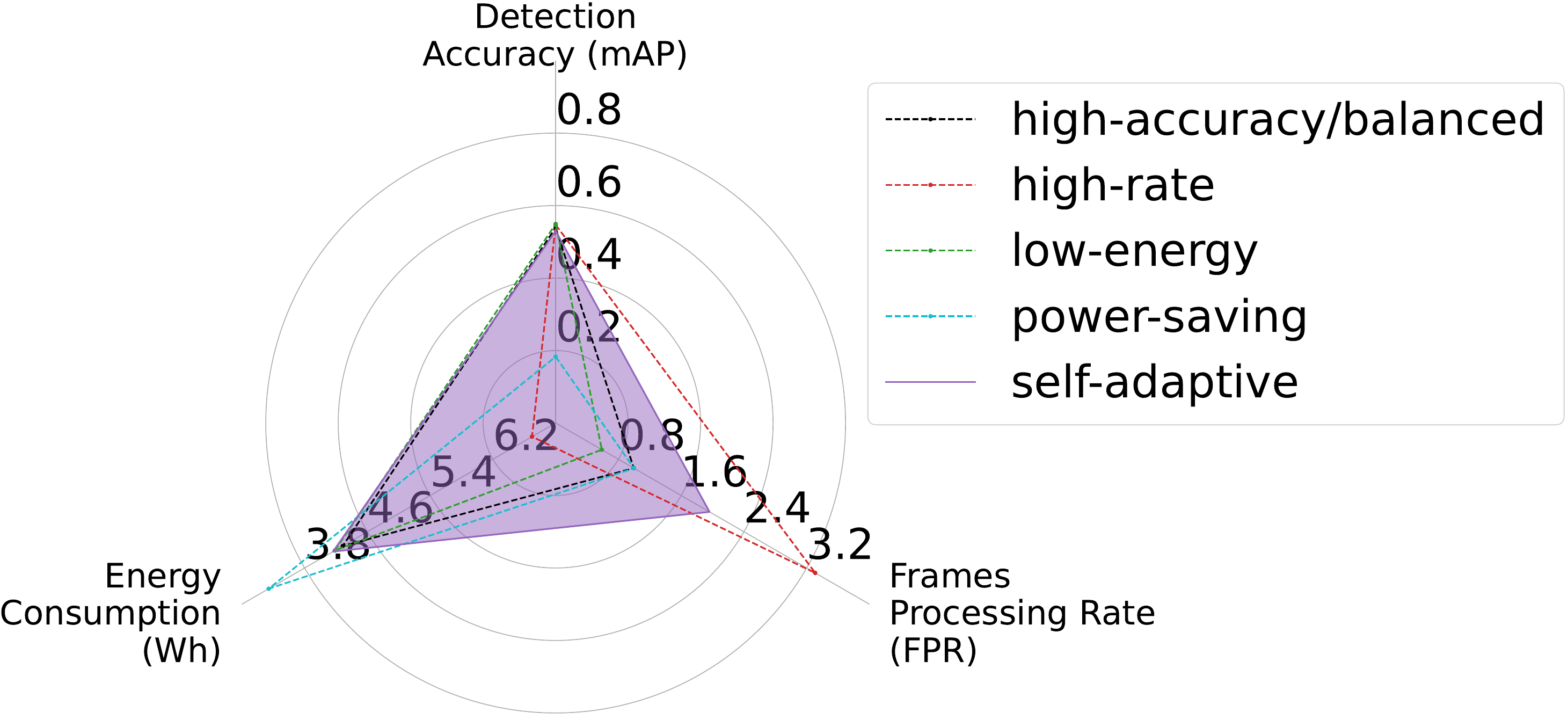}
         \caption{Weekdays Scenario}
         \label{fig:weekdays-radar}
     \end{subfigure}
     \begin{subfigure}[t]{0.8\columnwidth}
         \centering
         \includegraphics[width=\textwidth]{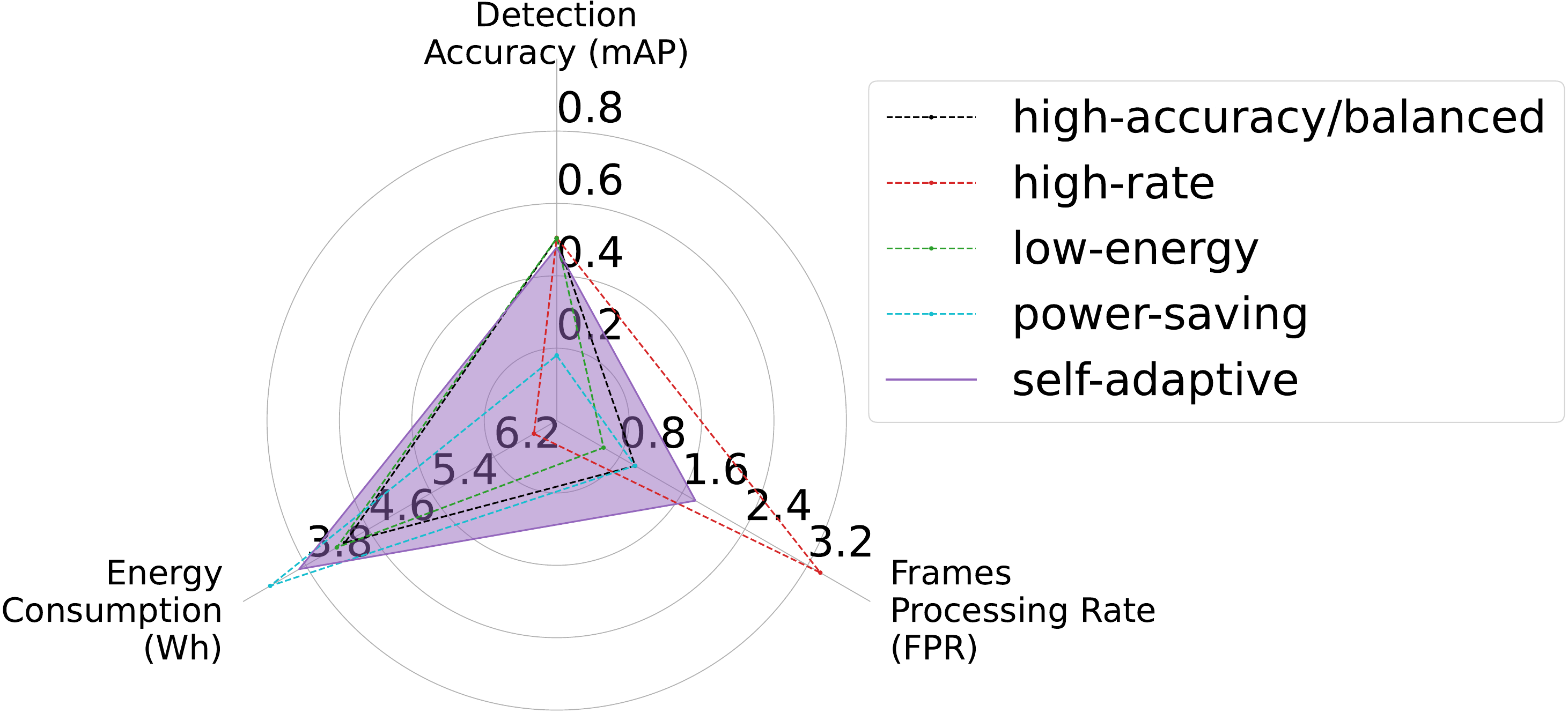}
         \caption{Weekends Scenario}
         \label{fig:weekends-radar}
     \end{subfigure}
     \caption{Radar charts comparing the SAA and the 4 non-adaptive applications in the weekdays and weekends scenarios.}
     \label{fig:radar-charts-comparison}
\end{figure}

Fig.~\ref{fig:radar-charts-comparison} compares the performance of the SAA (purple, solid line) with the four non-adaptive applications (black/red/green/cyan, dotted lines) in both the \emph{weekdays} (Fig.~\ref{fig:weekdays-radar}) and \emph{weekends} (Fig.~\ref{fig:weekends-radar}) scenarios. As for the radar charts in Fig.~\ref{fig:radars-heuristic-vs-exhaustive}, the higher the better.

The shape of the triangle in both the radar charts visually shows how the adaptive behavior guarantees the achievement of a better trade-off among the three objectives compared to the non-adaptive behavior. It outperforms three out of four non-adaptive applications regarding both energy consumption (i.e., \emph{\fewpedestriansmode}, \emph{\smallgroupmode}/\emph{balanced}, \emph{\crowdmode}) and FPR (i.e., \emph{\nopedestriansmode}, \emph{\fewpedestriansmode}, \emph{\smallgroupmode}/\emph{balanced}), and one out of four w.r.t. the detection accuracy (i.e., \emph{\nopedestriansmode}). Notably, it is still able to guarantee a similar accuracy when compared to the other three non-adaptive applications (i.e., \emph{\fewpedestriansmode}, \emph{\smallgroupmode}/\emph{balanced}, \emph{\crowdmode}).

In particular, compared to the best/worst non-adaptive operation mode, the SAA is able to save between 0.5\% and 61\% of energy in the \emph{weekdays} scenario, and between 13\% and 81\% in the \emph{weekends} scenario. The improvement on the FPR is between 96\% and 233\% in the \emph{weekdays} scenario, and between 77\% and 196\% in the \emph{weekends} scenario. The accuracy loss is between 2\% and 4\% in the \emph{weekdays} scenario, and between 5\% and 6\% in the \emph{weekends} scenario, but the SAA outperforms the \emph{\nopedestriansmode} application with a gain in the accuracy between 62\% and 189\%.

The SAA performed slightly differently in the two scenarios. In fact, the presence of a 3-peaks structure with a higher number of pedestrians in the \emph{weekdays} scenario makes the self-adaptive application to use more accurate and faster operation modes (i.e., \emph{\smallgroupmode} and \emph{\crowdmode}) for a larger amount of time, resulting in a higher FPR at the cost of a higher energy consumption.
On the other hand, the traffic shape of the \emph{weekends} scenario fosters the usage of energy efficient operation modes (i.e., \emph{\nopedestriansmode} and \emph{\fewpedestriansmode}), resulting in a lower energy consumption and slower processing speed. 
This shows how the SAA application can employ more accurate operation modes when the pedestrians workload is higher, using less accurate operation modes (i.e., \emph{\nopedestriansmode}) when the pedestrians workload is less demanding.

\begin{figure*}[!th]
     \centering
     \begin{subfigure}[t]{\textwidth}
         \centering
         \includegraphics[width=\textwidth]{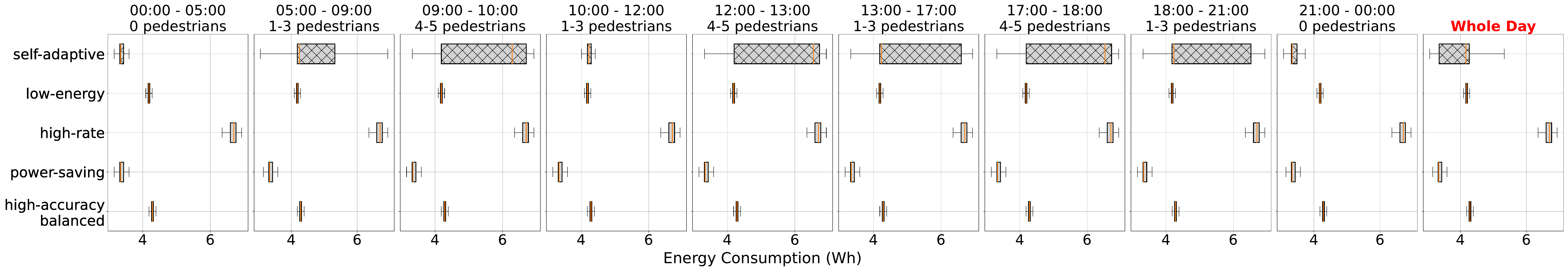}
         \caption{Weekdays Scenario}
         \label{fig:weekdays-power-boxplot}
     \end{subfigure}
     \hfill
     \begin{subfigure}[t]{\textwidth}
         \centering
         \includegraphics[width=\textwidth]{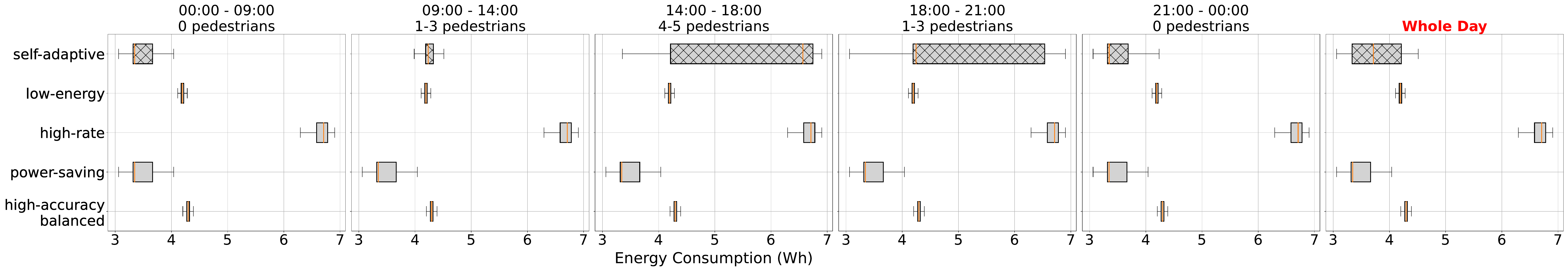}
         \caption{Weekends Scenario}
         \label{fig:weekends-power-boxplot}
     \end{subfigure}
     \caption{Box-plots comparing energy consumption for the self-adaptive and the four non-adaptive applications.}
     \label{fig:power-boxplot-comparison}
\end{figure*}

This behavior is also confirmed by the energy consumption box plots shown in Fig.~\ref{fig:weekdays-power-boxplot} and Fig.~\ref{fig:weekends-power-boxplot}. The two figures show the energy consumption of the self-adaptive application and the four non-adaptive applications in different time windows of the day for both the scenarios. The vertical orange line in the boxes indicates the median value.

We can observe how the self-adaptive application captures correctly the 3-peaks structure in the \emph{weekdays} scenario (Fig.~\ref{fig:weekdays-power-boxplot}) and uses the \emph{\crowdmode} in these three time windows. At the same time, it employs energy efficient operation modes (i.e., \emph{\nopedestriansmode} and \emph{\fewpedestriansmode}) when the pedestrians traffic is less intense (e.g., 00:00 - 05:00 and 21:00 - 00:00). A similar behavior is obtained in the \emph{weekends} scenario shown in Fig.~\ref{fig:weekends-power-boxplot}. In a nutshell, the self-adaptive solution is consuming energy only when it is worth doing it.

\medskip

In summary, the empirical evaluation shows how the proposed self-adaptive approach is capable of adapting to a changing environment while balancing multiple application requirements and energy consumption, behaving as optimally as the configurations selected with a near-exhaustive exploration of the parameters space. The experimental material to fully reproduce our study, including instructions to recreate our test-bed based on Raspberry Pi, is available at \repo.

\subsection{Threats to Validity}

\textit{First}, the design of the FSM requires the definition of a set of operations modes characterized by weights and thresholds, and the definition of state transition conditions. This is a manual \revised{MR.C1}{and non-trivial} operation guided by domain-expert knowledge that can \revised{MR.C1}{limit the feasibility of the approach and} lead to different results. Nevertheless, the reported results show how a SAA can largely outperform non-adaptive baselines, regardless of the specific configuration used. \textit{Second,} the design of the pedestrian  traffic shapes may have an impact on the results. To mitigate this threat, we referred to real scenarios to achieve realistic and informative results. 
\textit{Finally,} the results may not generalize to other application domains. Indeed, we proposed a \emph{case study} evaluation focusing on AI-services for pedestrian detection running at the edge, and the design of a SAA addressing a different problem may produce different results. Although the methodology and the approach are general, we cannot claim the results shall straightforwardly generalize to other contexts. The illustrated case study nevertheless provides evidence that the proposed approach can generate useful results in non-trivial domains such as pedestrian detection, which requires to balance high-speed computations (e.g., video-processing) with energy saving requirements. 



 
\section{Related Work}
\label{sec:related-work}

In the context of IoT architectures and edge oriented systems, self adaptation and optimization technologies have been used to address a range of aspects. 
For instance, adaptation capabilities have been engineered to achieve \emph{auto-scaling and task offloading}~\cite{alfonso2021modeling}, introducing flexibility in the computation at the cost of  some jitter in the quality of service and, often, not optimized energy consumption shifts among the nodes~\cite{jiang2020energy}.

Multiple approaches have been defined to modify the behavior of the components at the edge. The most common examples of self-adaptive edge components are those related to \emph{Adaptive Sampling}. Adaptive sampling refers to the idea of dynamically modifying the sampling rate of sensors and software monitoring probes as well as the inference rate of the components that process such data, according to the context~\cite{giouroukis2020survey,colombo2022towards, mertz2023software, yuan2012characterizing}. Collecting and transmitting less data can save energy and computational resources~\cite{jain2004adaptive}. 

Similarly, \emph{Adaptive Filtering} focuses on reducing the number of samples transmitted. For example, if a sensor value is considered similar to a previously collected value or evolves in a predicable way, a node can avoid the transmission of such information to save the transmission cost. Since  filtering usually results in sub-optimal performance, the filters must adapt at run-time to guarantee a consistent behavior~\cite{giouroukis2020survey}.

\emph{Adaptive Compression} has been also extensively exploited at the edge. Adaptive Compression solutions aim at reducing the data traffic in the network by reducing the size of the data packets with minimal loss, for instance using strategies that consider the importance of the processed data~\cite{lu2020adaptively}.
Different compression algorithms may also be used dynamically based on the shape of the data, enabling higher compression without inducing significant losses in the accuracy of the  data~\cite{cheng2012stcdg}.

The approach presented in this paper is complementary to all these forms of adaptation. In fact it provides a methodology to design a SAA running at the edge that is able to adapt its operation mode according to the context. The configurations that correspond to the operation modes are determined empirically, according to the key application objectives that must be optimized. Further optimizing sampling, filtering and compression strategies are  additional capabilities.

Self-adaptive behaviors to improve energy consumption have been also studied at the \emph{architectural level}~\cite{jiang2020energy}. 
For instance, a number of approaches have been proposed to target specific aspects of energy-awareness such as memory handling~\cite{jia2018maximum}, networking~\cite{baktir2017can}, storage~\cite{vales2019energy}, and scheduling and provisioning~\cite{aral2021ares}. Furthermore, the ever growing interest in machine learning based solutions lead to specific optimized models for the edge~\cite{brandic2021sustainable}. These solutions can address specific dimensions but lack both the state-based adaptation capabilities introduced in this paper, and the definition of a practical empirical procedure to determine the concrete configurations that must be used by the SAA. \revised{MR.C4, R1.C6, R2.C5}{Conversely, Da Silva et al.~\cite{da2011framework} proposed a framework for the automatic generation of application processes. Such processes represent the goals and capabilities of the application in the form of application workflows. This level of adaptation is not usually suitable for edge applications, since the run-time generation of the application processes requires extensive computational capabilities and introduces significant computational overhead~\cite{camara2015optimal}, which may not be available at edge.

\emph{Mobile applications} is another domain of self-adaptation where energy consumption is pivotal~\cite{grua2019self}. While adaptation mechanisms designed for mobile applications are not directly comparable to applications running on the Edge, they share some key aspects, such as the presence of a resource-constrained and battery-powered devices. For instance, Ardito et al.~\cite{ardito2013energy, ardito2013glcb} proposed an architectural paradigm in which the operating system or the middleware is able to offer energy-related information to running applications. This enables the implementation of energy-aware self-adaptation strategies based on energy levels. Our proposal is orthogonal with respect to this approach, as we investigate how to design and deploy such applications, with specific focus on those that are AI-based, but without assuming run-time information about the available energy.}



 
\section{Conclusions}
\label{sec:conclusions}

We presented an approach that can guide developers in the implementation of AI-based self-adaptive applications able of switching their operation modes in response to changes in the environment. The configuration of the operation modes are determined empirically, based on  a meta-heuristic search procedure that can identify useful configurations by sampling a small portion of the configuration space. Experimental results show how the proposed approach can outperform non-adaptive baseline configurations, behaving as optimally as configurations selected with a nearly exhaustive exploration of the configuration space, in a pedestrian detection scenario.

Future work concerns with \revised{R2.C1}{automating the FSM design and synthesis through data-driven methods, and} extending the self-adaptive capabilities by considering clusters of instances that can adapt simultaneously. We also plan to study our approach in a more complex setup involving battery-powered devices and photovoltaic panels, considering run-time energy-related metrics and deploying our prototype in the field.

\section*{Acknowledgments}


This work has been partially supported by the MUR under the grant ``Dipartimenti di Eccellenza 2023-2027", Engineered MachinE Learning-intensive IoT systems (EMELIOT) national research project which has been funded by the MUR under the PRIN 2020 program (Contract 2020W3A5FY), Runtime Control in Multi Clouds (RUCON), Austrian Science Fund (FWF): Y904-N31 START-Programm, 2015, Sustainable Watershed Management Through IoT-Driven Artificial Intelligence (SWAIN), CHIST-ERA-19-CES-005, Austrian Science Fund (FWF), 2021, Standalone Project Transprecise Edge Computing (Triton), Austrian Science Fund (FWF): P 36870-N, 2023, Flagship Project High-Performance Integrated Quantum Computing (HPQC) \# 897481 Austrian Research Promotion Agency (FFG), 2023, and by the 5G Use Case Challenge InTraSafEd 5G (Increasing Traffic Safety with Edge and 5G) funded by the City of Vienna.

\bibliographystyle{IEEEtran}
\bibliography{references}

\end{document}